# Myelin Distribution at the Optic Nerve Myelination Transition Zone Influences Axonal Biomechanics

Tingting Liu[1], Xiaofei Wang[2,3], C. Ross Ethier[4,5,6], Martin Buist[7], Tin Aung[1,8], Michaël J. A. Girard[1,4,5,8,9]

1. Singapore Eye Research Institute, Singapore National Eye Centre, Singapore
2. Beijing Advanced Innovation Center for Biomedical Engineering, Key Laboratory for Biomechanics and Mechanobiology of Ministry of Education, School of Biological Science and Medical Engineering, Beihang University, Beijing, China
3. Beijing Visual Science and Translational Eye Research Institute (BERI), Beijing Tsinghua Changgung Hospital, Tsinghua Medicine, Tsinghua University, Beijing, China
4. Department of Biomedical Engineering, Georgia Institute of Technology/Emory University, Atlanta, GA, United States
5. Department of Ophthalmology, Emory University, Atlanta, GA, United States
6. School of Mechanical Engineering, Georgia Institute of Technology, Atlanta, GA, United States
7. Department of Biomedical Engineering, National University of Singapore, Singapore
8. Duke-NUS Graduate Medical School, Singapore
9. Emory Empathetic AI for Health Institute, Emory University, Atlanta, GA, United States

**Word count:** 3,830 (Manuscript Text)
249 (Abstract)

**Tables:** 1
**Figures:** 7

**Conflict of Interest:** MJAG is the co-founder of the AI start-up company Abyss Processing Pte Ltd

**Corresponding Author:** Michaël J. A. Girard
Ophthalmic Engineering & Innovation Laboratory (OEIL)
Emory Eye Center, Emory School of Medicine
Emory Clinic Building B, 1365B Clifton Road, NE
Atlanta GA 30322
mgirard@ophthalmic.engineering

# ABSTRACT

**Purpose:** The lamina cribrosa (LC) is considered the initial site of glaucomatous retinal ganglion cell (RGC) injury, and is also the region where unmyelinated RGC axons become myelinated. Here we sought to use finite element (FE) modeling to investigate how the configuration of the myelination transition zone (MTZ) influences the mechanical insult to RGC axons.

**Methods**: A multiscale FE framework was developed to investigate the biomechanical effect of myelin distribution on IOP-induced axonal stress and strain at the MTZ. An anatomically based macro-scale FE eye model was used to compute LC deformations under 15 and 45 mmHg IOP. These deformations were then applied to micro-scale models of the posterior LC, consisting of axons, myelin sheaths, and surrounding matrix. Four distinct MTZ boundary configurations were simulated: one flat and three with random posterior offsets of 3, 6, or 9 μm, representing potential physiological variations. IOP-induced effective axonal strains and stresses were quantified across the different MTZ configurations.

**Results**: Under IOP loading, axons exhibited longitudinal compression and transverse stretch, with marked effective stress and strain discontinuities at the myelin boundary. Across all models, the unmyelinated region exhibited higher effective stress and strain than the myelinated region, and this mechanical discontinuity increased with larger MTZ offsets.

**Conclusions**: Glaucoma-associated demyelination has been previously suggested to precede RNFL thinning. Here we have shown that the MTZ configuration directly influences RGC axonal mechanics. Whether different MTZ profiles can initiate glaucomatous injury, whether demyelination accelerates disease progression, or whether both mechanisms contribute, remains to be determined.

# INTRODUCTION

Glaucoma, the leading cause of irreversible blindness, is characterized by progressive damage to retinal ganglion cell (RGC) axons, optic nerve atrophy and visual field loss.[1] The lamina cribrosa (LC), where RGC axons exit the eye within the optic nerve head (ONH), is considered the primary site of glaucomatous injury.[2] Notably, as RGC axons traverse the LC and its immediate retrobulbar region, they undergo a transition from an unmyelinated to a myelinated state, in a region denoted as the myelination transition zone (MTZ).[3]

The MTZ is not a discrete anatomical boundary but instead forms a gradual and spatially heterogeneous interface, with myelin sheaths extending anteriorly into the posterior LC to variable depths (**Figure 1)**. Myelin is primarily recognized for providing electrical insulation and supporting rapid axonal conduction; however, it also contributes to the structural support, radial constraint, and local mechanical properties of axons.[4] The presence of myelin markedly increases local stiffness relative to adjacent unmyelinated regions.[5,6] Consistent with this effect, differences in myelin content contribute to distinct mechanical properties between white matter–gray matter interface in the brain[7,8]. More broadly, studies of biological transition zones, such as the tendon-bone enthesis[9,10], have shown that these material mismatches are known to influence local mechanical behavior.

This biomechanical transition may be particularly relevant to glaucoma because the MTZ lies immediately posterior to the LC, the principal site of IOP-related axonal injury. Deformation of the LC and retrolaminar tissues may therefore be transmitted across axonal regions with different myelination states, potentially concentrating mechanical insults near the myelin boundary. Moreover, animal studies have demonstrated early alterations in myelin integrity that precede measurable retinal nerve fiber layer thinning and which occur prior to neuronal death and axonal dysfunction.[11] Collectively, these findings highlight the potential role of myelin distribution in modulating axonal vulnerability, and emerging evidence from animal models suggests that myelin disruption may serve as an early biomarker of glaucomatous damage.[12]

Despite its potential importance, the biomechanical role of myelin and the biomechanical influence of the MTZ remain poorly understood. Previous biomechanical studies using the finite element (FE) method have mostly focused on macroscopic LC-level behavior[13–16], modeling the LC as a continuum without including axons or myelin. At the microscale, other studies have examined axonal mechanics focusing on axonal deformation in neural tissues outside the ONH, without explicitly accounting for the mechanical contribution of myelin.[17–19] Similarly, in vivo investigations have centered on RGC axons without explicitly considering the biomechanical role of myelin.[20,21] Consequently, it remains unclear how the geometric configuration of myelin in the MTZ influences the stress–strain environment of RGC axons under normal conditions, and whether specific MTZ configurations exacerbate axonal vulnerability under elevated IOP in glaucoma. Addressing this question may help establish a biomechanical link between early myelin alterations and axonal vulnerability in glaucoma.

The aim of this study was to use FE modeling to investigate how the configuration of the MTZ influences IOP-induced stress and strain in axons, thereby exploring potential biomechanical mechanisms that may contribute to axonal vulnerability in glaucoma.

# METHODS

A multiscale FE framework was developed to investigate the biomechanical effect of myelin distribution on IOP-induced axonal stress and strain at the MTZ. At the macroscale (millimeters), we simulated ONH deformations that can be observed clinically, such as LC displacement and strain. At the microscale (micrometers), we focused on the local mechanical environment of individual axons and surrounding myelin within the MTZ, which cannot be directly measured in vivo. This multiscale approach allows microscale stresses and strains to be estimated based on the macro-scale ONH deformations, thereby connecting clinically observable ONH changes to the microscale mechanical factors that may contribute to axonal vulnerability in glaucoma.

Here, the macro-scale FE model (**Figure 2A**) represented the whole eye globe and was used to compute LC deformations under a normal IOP of 15 mmHg and an elevated IOP of 45 mmHg, based on our previous work.[16,22] These deformations were then applied to micro-scale models localized at the MTZ of the posterior LC (**Figure 2B**), consisting of axons, with myelin sheaths, and surrounding matrix. To evaluate the effect of myelin distribution, four distinct MTZ boundary configurations were simulated, incorporating posterior offsets of 0, 3, 6, and 9 μm. These spaced offsets were selected as exploratory parametric variations within the dimensions of the micro-scale model. Finally, IOP-induced effective axonal strains and stresses were quantified across different MTZ configurations. Effective strain and stress are scalar measures derived from the corresponding strain and stress tensors and were used to characterize the overall magnitude of the mechanical response within the axons.

## Geometry of the FE Models

**Macro-scale Geometry.** The macro-scale (eye globe) model, including its tissue components and geometric dimensions, was consistent with our previous studies.[15,16,22,23] It included the sclera, choroid, retina, LC, ON, pia mater, dura mater, orbital fat-muscle complex (OFM) and orbital wall. As in these previous models, the eye globe was modeled as a sphere (outer diameter, 24 mm), with the ON having a circular coronal cross-section (diameter, 3.0 mm). The ONH and surrounding tissues were assigned literature-based average dimensions: scleral thickness (perilimbal 0.8 mm, equatorial 0.5 mm, posterior 1.0 mm), choroid 0.16 mm, retina (perilimbal 0.1 mm, posterior 0.32 mm), Bruch's membrane 0.005 mm, LC radius 1.0 mm, LC thickness 0.3 mm, and cup depth 0.33 mm. The ON length from the ONH to the orbital apex was 24.8 mm, with pia and dura thickness of 0.0612 and 0.313 mm, respectively; dural thickness increased at the scleral flange, as observed histologically.[24] Extraocular muscles and orbital fat were modeled as a single entity. Only half of the eye was constructed, assuming symmetry about a transverse plane through the globe center (**Figure 2A**). The model was meshed with 75,290 nodes and 68,472 elements, including 63,000 8-node hexahedra, 2,412 6-node pentahedra and 3,060 4-node shell elements.

**Micro-scale Geometry.** An idealized micro-scale geometry was defined within a macro-scale element located in the posterior portion of the LC, which was selected to represent the local mechanical environment of the MTZ. The same micro-scale geometry was used for all simulations, with overall dimensions of 85 × 30 × 94 μm³. It comprised 160 longitudinally oriented axons (approximately 50% of a fiber bundle[25–27]) plus their myelin sheaths, and the surrounding matrix. Axon radii were randomly assigned following a Log-normal distribution (Mean: 1.00 μm; Median: 0.77 μm; SD: 0.06 μm; Range: 0.1–1.5 μm).[27] Myelin thickness was determined using a g-ratio of 0.7, [28] defined as the ratio of the axon radius to the outer radius of the myelin sheath, with the interstitial space filled by the matrix. Four distinct MTZ boundary configurations were simulated: one flat reference configuration, in which the myelin height was set to half of the axon height, and three configurations with posterior offsets randomly distributed within 0-3, 0-6, or 0-9 μm from the reference MTZ boundary (**Figure 2C-F**).

The geometric configurations and meshing process were automated using the Gmsh Python API [29]. This allowed for automated parametric modeling in which axon radii and spatial distributions were sampled from the prescribed Log-normal distribution. Conformal meshing at the interfaces between axon and myelin was achieved via OpenCASCADE boolean fragments, ensuring that adjacent regions shared common nodes across material boundaries, thereby avoiding mesh discontinuities. The four micro-scale models were meshed using a prescribed element size of 0.8 μm, which resulted in 506,967 nodes and 2,984,041 three-dimensional linear tetrahedral elements for the flat reference model, and 504,602 nodes and 2,971,046 elements, 500,651 nodes and 2,944,413 elements, and 498,458 nodes and 2,931,182 elements for the 3, 6, and 9 μm posterior offset models, respectively. A mesh analysis confirmed that the regional mean micro-scale stress and strain outcomes were insensitive to mesh resolution, with less than 1% variation across the evaluated mesh densities (**Supplementary Material A**).

## Macro to Micro Coupling and Spatial Sensitivity Analysis

To assess the spatial sensitivity of the micro-scale predictions, five macro-scale elements were selected from the posterior element layer of the LC (**Figure 2B**). Three elements were selected along a common radial transect, representing central, intermediate, and peripheral LC locations. Two additional elements at comparable radial positions represented the nasal and temporal regions to evaluate circumferential variation.

For each location and IOP condition, the corresponding deformation gradient tensor was extracted from the macro-scale model and applied separately to the same micro-scale geometry for all four MTZ configurations. The micro-scale geometry and microstructural arrangement were held constant across locations, thereby isolating the influence of spatial variation in the macro-scale loading environment. The macro-scale element used to define the dimensions of the micro-scale geometry was designated as the primary location. Results from the primary location were used for the primary presentation of the stress and strain distributions, while all five locations were included in the spatial sensitivity analysis. The selected locations and corresponding deformation gradient tensors are provided in **Supplementary Material B**.

## Biomechanical Properties of Macro- and Micro-Components

Tissue biomechanical properties were assigned based on values reported in previous studies of human ocular tissues. For the macro-model, the biomechanical properties of ocular tissues were the same as those used in our previous studies.[15,16,22,23] Specifically, the sclera and LC were modeled as fiber-reinforced composites (Mooney-Rivlin Von Mises Distributed Fibers) and all other tissues were modeled as “isotropic elastic” materials as in **Table 1**.

For the micro-scale model, definitive biomechanical properties for individual axons and myelin are poorly characterized. Therefore, biomechanical properties were derived from available comparative neuro-biomechanical data. Previous study have demonstrated that brain stiffness increases with myelin content.[5] This is further supported by atomic force microscopy measurements, which reported significantly higher elastic modulus values for myelinated axonal segments (3,790 ± 1,179 Pa) compared to unmyelinated segments (3,411 ± 1,316 Pa; $p < 0.0001$).[6]

To determine the stiffness of the myelin sheath, we applied a weighted composite model based on the above experimental measurements. Specifically, the elastic modulus of the unmyelinated axon segment (3,411 Pa) was used as the internal axon stiffness. Based on a g-ratio of 0.7, [28] the volume fractions for the axon ($Va = g^2$) and the myelin ($Vm = 1 - g^2$) were determined. Using the rule of mixtures for the total measured modulus ($E = Em * Vm + Ea * Va$), the stiffness of myelin was calculated to be 4,154 Pa, corresponding to approximately 1.22 times that of axons. Furthermore, as axons have been reported to be approximately three times stiffer than the surrounding matrix,[30] the elastic modulus value for the matrix was set to 1,137 Pa (**Table 1**). All micro-scale constituents in the model were assumed to be incompressible.

## Contacts, Boundary, and Loading Conditions

For the macro-scale model, contact definitions and boundary conditions followed our previous studies.[15,23] Briefly, the OFM and the dura were tied together, the OFM could slide over the orbital margin, and the corneo-scleral shell and the OFM had frictionless sliding contact. All other tissues of the eye were bonded together by sharing nodes at their boundaries. Nodes on the symmetry plane were constrained in the normal direction to maintain symmetry. The fat and ON were fixed at the orbital apex to simulate their fibrous attachment to the bony optic nerve canal. Additionally, the outer boundary of the orbital bone was fixed to represent its rigid support. Two loading conditions were considered: a baseline IOP of 15 mmHg[31] and an elevated IOP of 45 mmHg.[32] For both cases, the cerebrospinal fluid pressure (CSFP) was set to 12.9 mmHg.[33]

For the micro-scale model, a multiscale framework was employed to transfer macro-model deformation to the micro-model. Specifically, the deformation gradients of the macro-model element corresponding to the selected micro-model region was extracted under baseline and elevated IOP loading conditions, and applied as boundary conditions to drive the micro-scale simulations. These boundary conditions satisfy the Hill–Mandel macro–micro energy equivalence,[34] ensuring that the volume-averaged microscale strain corresponds to the macroscopic strain. All geometric and displacement quantities were expressed in micrometer units to maintain consistency across scales.

## FE Processing to Estimate IOP-induced Axonal Stresses and Strains

In total, macro-scale FE simulations were performed under baseline and elevated IOP conditions. These macro-scale simulations provided boundary conditions for corresponding micro-scale models, representing four distinct

MTZ configurations, each evaluated under both baseline and elevated IOP conditions.

For the micro-scale model, effective stresses and strains of individual axons were quantified within defined regions different MTZ configurations. Two regions were defined relative to the MTZ interface of each axon (**Figure 3**): (1) an unmyelinated region, extending from the MTZ boundary to 3 μm anterior to the MTZ boundary (toward the LC), and (2) a myelinated region, extending 3 μm posterior to the MTZ boundary (toward the retrolaminar ON). Axonal effective stress and strain were spatially averaged within each region, and differences between these two regions were calculated to quantify the mechanical discontinuity across the myelin boundary.

All models were solved using FEBio (Version 4.0; Musculoskeletal Research Laboratories, University of Utah, Salt Lake City, UT, USA).[35] All data processing and statistical analyses were performed using Python (version 3.13.5; Python Software Foundation, Wilmington, DE, USA).

# RESULTS

## Macro-Scale Deformation Under Normal and Elevated IOP

Relative to the same unloaded reference state, the mean effective strain of the entire LC increased from 0.82% under normal IOP to 3.09% under elevated IOP. At the primary location, corresponding to the macro-scale element used to define the dimensions of the micro-scale geometry, the effective strain increased from 0.96% under normal IOP to 4.77% under elevated IOP.

Under normal IOP, the selected element exhibited mild longitudinal compression in the axonal direction (0.52%) with slight transverse stretch perpendicular to the axons (0.26%). Under elevated IOP, this deformation pattern became more pronounced, with 2.56% longitudinal compression and 1.26% transverse stretch. The same overall deformation mode was observed at other four locations, although the deformation magnitudes varied. The full deformation gradient tensors applied to the micro-scale models are provided in **Supplementary Material B**.

## Stress and Strain Discontinuities Across the MTZ

Under elevated IOP loading, a clear discontinuity in both axonal stress and strain was observed at the MTZ for all MTZ configuration models (**Figure 4).** A similar but less pronounced pattern was also observed under normal IOP. Although no consistent qualitative stress or strain pattern was observed across different myelin offset models, quantitative regional analysis demonstrated a consistent difference between the unmyelinated and myelinated regions.

Under both normal and elevated IOP conditions, the unmyelinated region exhibited higher mean effective stresses and strains than the myelinated region across all models (**Figure 5**). At the primary location, the stress and strain differences across the MTZ showed an overall increase with larger myelin offsets.

At the single-axon level, axonal stress and strain exhibited abrupt changes across the MTZ boundary, confirming the mechanical discontinuity at the MTZ (**Figure 6**).

## Spatial Sensitivity of the MTZ Mechanical Discontinuity

Across all five element locations, the unmyelinated region consistently exhibited higher effective stress and strain than the myelinated region under both normal and elevated IOP conditions, indicating that the direction of the mechanical discontinuity across the MTZ was spatially robust (**Figure 7**).

Under normal IOP, the stress and strain differences across the MTZ generally increased with greater myelin offset at all five locations, although some responses approached a plateau at the larger offsets. Under elevated IOP, the same overall increasing trend was observed at the center, intermediate, nasal, and temporal locations. In contrast, at the peripheral location, the stress difference progressively decreased with increasing offset, while the strain difference remained relatively stable.

Along the radial transect, the mechanical discontinuity was greatest at the central location and smallest at the peripheral location, particularly under elevated IOP. For the 9-μm offset configuration under elevated IOP, the stress difference ranged from 1.21 Pa at the peripheral location to 7.60 Pa at the central location, while the corresponding strain difference ranged from 0.06% to 0.49%. The nasal and temporal locations exhibited the same overall offset-dependent direction, although the temporal location showed a greater increase with myelin offset. Detailed results are provided in **Supplementary Material C**.

# DISCUSSION

This study used a multiscale FE framework to investigate how myelin distribution at the MTZ modulates IOP-induced axonal biomechanics. Under IOP loading, axons exhibited localized stress and strain discontinuities across the myelin boundary. Larger MTZ offsets generally amplified the mechanical discontinuity across the MTZ, and this effect depended on the local macro-scale loading environment. However, the magnitude of these changes was small, and their physiological significance remains to be determined.

We observed longitudinal compression and transverse stretch of axons under elevated IOP loading. Mechanically, longitudinal compression refers to shortening of the axon along its long axis, whereas transverse stretch reflects tensile expansion perpendicular to the axonal axis. Because mild longitudinal compression and transverse stretch were also present under normal IOP, this deformation pattern should not be interpreted as inherently pathological. Our results suggest that elevated IOP amplifies this existing deformation, which may become mechanically unfavorable when the magnitude or duration of deformation exceeds the tolerance of axonal structures. Axonal function depends on the integrity and spatial organization of the axonal cytoskeleton. Microtubules and neurofilaments are predominantly aligned along the axonal axis and provide structural support, while microtubules also serve as essential tracks for axonal transport.[36] Therefore, excessive longitudinal compression may promote local cytoskeletal bending or buckling,[37] thereby disrupting cytoskeletal alignment and potentially impairing axonal transport, whereas transverse stretch may alter the cross-sectional organization of the axon and increase mechanical mismatch between the axolemma, cytoskeleton, and surrounding myelin, potentially compromising axonal integrity.[38] Although the present model does not directly simulate cytoskeletal failure or transport dysfunction, the predicted deformation pattern is consistent with previous computational studies showing that axons within the LC and post-LC regions experience longitudinal compression and transverse stretch under IOP elevation.[39] In addition, impaired axonal transport at the ONH is considered an early event in RGCs degeneration in glaucoma.[4,40,41] These findings provide mechanistic and pathological context, but the physiological significance of the predicted deformation magnitudes remains to be established.

Regional analysis demonstrated higher effective stress and strain in the unmyelinated region than in the myelinated region across all MTZ configurations. However, its magnitude and response to myelin offset were spatially heterogeneous. The discontinuity was greater at the center than at the peripheral location, particularly under elevated IOP, and the nasal and temporal locations differed in magnitude despite exhibiting the same overall directional trend. Thus, the mechanical consequences of altered myelin distribution depend not only on MTZ configuration but also on location within the heterogeneous LC deformation field. However, these changes were modest, and their physiological significance remains uncertain. Whether such localized mechanical differences contribute to axonal injury, or demyelination during glaucoma progression require further investigation.

At the single-axon level, the elevations of stress and strain at the myelin boundary are likely driven by the abrupt transition in material properties and geometric constraints between myelinated and unmyelinated segments. A previous study has reported similar findings, showing local stress amplification near the gray–white matter interface.[8] This discontinuity alters load transfer and gives rise to stress concentration, analogous to interface effects observed in composite materials and other biological transition zones.[42] To minimize the influence of axon size and spatial location, axons at the same spatial locations with same radii were selected for comparison among the four models. Among these matched axons, a larger MTZ offset was associated with greater stress discontinuity at the myelin boundary when differences in myelin depth were relatively small (**Supplementary Material D**).

The MTZ is a biologically distinct transition between unmyelinated axons within the ONH and myelinated axons in the retrolaminar optic nerve. One study identified the MTZ using myelin basic protein immunolabeling and reported axonal transport-related protein accumulation in both the unmyelinated ONH and myelinated optic nerve after IOP elevation, suggesting that axonal transport impairment occurs in regions surrounding the unmyelinated-to-myelinated transition.[43] Recent nonhuman primate experimental glaucoma studies have shown early retrolaminar myelin disruption and demyelination in structurally intact optic nerve axons, indicating that myelin disruption may occur before overt axonal degeneration.[44,45] These observations provide biological motivation for examining whether variation in myelin distribution alters the local axonal mechanical environment. In the present models, a larger myelin offset or subtle disruption of myelin configuration could increase the mechanical discontinuity at the MTZ, amplifying the local stress and strain experienced by RGC axons. These findings raise the possibility that MTZ configuration may serve as potential biomechanical marker associated with regional axonal susceptibility in glaucoma. However, whether different MTZ profiles can initiate glaucomatous injury, whether demyelination accelerates disease progression, or whether both mechanisms contribute, remains to be determined.

***Limitations.*** This study is subject to several limitations. First, definitive material properties for individual axons and myelin are not yet well established. Accordingly, material parameters were adopted from available neuro-biomechanical literature and comparative studies[5,6,30], which may influence the absolute magnitude of the predicted stress and strain, but are unlikely to alter the observed relative trends. Second, the micro-scale model exhibited a boundary effect, characterized by localized stress and strain concentrations confined to a narrow (~5 μm) region near the external boundaries along the axonal axis. This effect primarily arises from artificial constraints imposed at the macro-scale element boundaries and local stiffness mismatches. As described in the methods, our analyses were located in the MTZ central region, where stress and strain distributions remained stable and insensitive to boundary conditions. Given that the boundary layer is small relative to the overall model height, it did not influence the main conclusions of this study. Third, the myelin offset amplitudes examined in this exploratory parametric analysis (3-9 μm) were substantially smaller than the extent of retrolaminar demyelination reported in experimental glaucoma studies[44,45] , in which myelin disruption and posterior shifts of myelin onset may extend from tens to several hundred microns posterior to the LC. Therefore, the present models should be interpreted as a simplified and localized biomechanical approximation of MTZ configurations, rather than a direct anatomical replication of glaucomatous retrolaminar demyelination. Fourth, the present models were not intended to identify a specific physiological myelin configuration that initiates damage, but rather to isolate the mechanical consequences of varying MTZ geometries under controlled loading conditions. Nonetheless, by systematically linking MTZ geometry to axonal mechanical exposure across IOP levels, this study provides mechanistic insight into how structural features at the MTZ may influence axonal vulnerability and offers a foundation for future studies integrating imaging, biology, and longitudinal disease modeling. Finally, the biological importance of myelin alterations in glaucoma pathogenesis remains uncertain. Although our simulations suggest that changes in myelin distribution can modify the local mechanical environment of RGC axons, the present study cannot determine whether myelin disruption contributes to the initiation or progression of glaucomatous damage. Experimental and longitudinal studies are needed to clarify the causal role and biological significance of myelin changes in glaucoma.

***Conclusions.*** In this study, we used multiscale FE models to study the biomechanical effect of myelin distribution on IOP-induced axonal stress and strain at the MTZ. The results demonstrated that MTZ configuration can influence local axonal biomechanics, while increasing MTZ offset amplitude enlarged the mechanical discontinuity across the MTZ. These findings raise the possibility that MTZ configuration may serve as a potential biomechanical marker associated with regional axonal susceptibility in glaucoma.

# ACKNOWLEDGMENTS

We acknowledge funding from (1) the Emory Eye Center (Emory University School of Medicine, Start-up funds, MJAG & CRE), (2) a Challenge Grant from Research to Prevent Blindness, Inc. to the Department of Ophthalmology at Emory University, (3) NIH grant P30EY06360 to the Atlanta Vision Research Community, (4) the National Eye Institute of the National Institutes of Health under award numbers R01EY037299 and R01EY037245 (MJAG), (5) the NMRC-LCG grant 'TAckling & Reducing Glaucoma Blindness with Emerging Technologies (TARGET)', award ID: MOH-OFLCG21jun-0003 (MJAG), and (6) the National Natural Science Foundation of China (12472304).

# REFERENCES

1. Kapetanakis VV, Chan MPY, Foster PJ, Cook DG, Owen CG, Rudnicka AR. Global variations and time trends in the prevalence of primary open angle glaucoma (POAG): a systematic review and meta-analysis. *Br J Ophthalmol*. 2016;100(1):86-93. doi:10.1136/bjophthalmol-2015-307223
2. Burgoyne CF, Crawford Downs J, Bellezza AJ, Francis Suh JK, Hart RT. The optic nerve head as a biomechanical structure: a new paradigm for understanding the role of IOP-related stress and strain in the pathophysiology of glaucomatous optic nerve head damage. *Progress in Retinal and Eye Research*. 2005;24(1):39-73. doi:10.1016/j.preteyeres.2004.06.001
3. Calkins DJ. Critical pathogenic events underlying progression of neurodegeneration in glaucoma. *Progress in Retinal and Eye Research*. 2012;31(6):702-719. doi:10.1016/j.preteyeres.2012.07.001
4. Dias MS, Luo X, Ribas VT, Petrs-Silva H, Koch JC. The Role of Axonal Transport in Glaucoma. *IJMS*. 2022;23(7):3935. doi:10.3390/ijms23073935
5. Weickenmeier J, De Rooij R, Budday S, Steinmann P, Ovaert TC, Kuhl E. Brain stiffness increases with myelin content. *Acta Biomaterialia*. 2016;42:265-272. doi:10.1016/j.actbio.2016.07.040
6. Chuang Y, Alcantara A, Fabris G, et al. Myelination dictates axonal viscoelasticity. *Eur J of Neuroscience*. 2023;57(8):1225-1240. doi:10.1111/ejn.15954
7. Sparrey CJ, Manley GT, Keaveny TM. Effects of White, Grey, and Pia Mater Properties on Tissue Level Stresses and Strains in the Compressed Spinal Cord. *Journal of Neurotrauma*. 2009;26(4):585-595. doi:10.1089/neu.2008.0654
8. Alisafaei F, Gong Z, Johnson VE, Dollé JP, Smith DH, Shenoy VB. Mechanisms of Local Stress Amplification in Axons near the Gray-White Matter Interface. *Biophysical Journal*. 2020;119(7):1290-1300. doi:10.1016/j.bpj.2020.08.024

9. Liu Y, Birman V, Chen C, Thomopoulos S, Genin GM. Mechanisms of Bimaterial Attachment at the Interface of Tendon to Bone. *Journal of Engineering Materials and Technology*. 2011;133(1):011006. doi:10.1115/1.4002641
10. Liu Y, Schwartz AG, Birman V, Thomopoulos S, Genin GM. Stress amplification during development of the tendon-to-bone attachment. *Biomech Model Mechanobiol*. 2014;13(5):973-983. doi:10.1007/s10237-013-0548-2
11. You Y, Joseph C, Wang C, et al. Demyelination precedes axonal loss in the transneuronal spread of human neurodegenerative disease. *Brain*. 2019;142(2):426-442. doi:10.1093/brain/awy338
12. Xue J, Zhu Y, Liu Z, et al. Demyelination of the Optic Nerve: An Underlying Factor in Glaucoma? *Front Aging Neurosci*. 2021;13:701322. doi:10.3389/fnagi.2021.701322
13. Sigal IA, Flanagan JG, Ethier CR. Factors Influencing Optic Nerve Head Biomechanics. *Invest Ophthalmol Vis Sci*. 2005;46(11):4189. doi:10.1167/iovs.05-0541
14. Hua Y, Voorhees AP, Sigal IA. Cerebrospinal Fluid Pressure: Revisiting Factors Influencing Optic Nerve Head Biomechanics. *Invest Ophthalmol Vis Sci*. 2018;59(1):154. doi:10.1167/iovs.17-22488
15. Wang X, Rumpel H, Lim WEH, et al. Finite Element Analysis Predicts Large Optic Nerve Head Strains During Horizontal Eye Movements. *Invest Ophthalmol Vis Sci*. 2016;57(6):2452. doi:10.1167/iovs.15-18986
16. Liu T, Wang K, Wang YX, et al. Ciliary muscle traction during accommodation is able to induce optic nerve head deformation. *Eye*. Published online January 4, 2025. doi:10.1038/s41433-024-03569-1
17. Karami G, Grundman N, Abolfathi N, Naik A, Ziejewski M. A micromechanical hyperelastic modeling of brain white matter under large deformation. *Journal of the Mechanical Behavior of Biomedical Materials*. 2009;2(3):243-254. doi:10.1016/j.jmbbm.2008.08.003
18. Montanino A, Kleiven S. Utilizing a Structural Mechanics Approach to Assess the Primary Effects of Injury Loads Onto the Axon and Its Components. *Front Neurol*. 2018;9:643. doi:10.3389/fneur.2018.00643

19. Wang X, Neely AJ, McIlwaine GG, Lueck CJ. Multi-scale analysis of optic chiasmal compression by finite element modelling. *Journal of Biomechanics*. 2014;47(10):2292-2299. doi:10.1016/j.jbiomech.2014.04.040
20. Knöferle J, Koch JC, Ostendorf T, et al. Mechanisms of acute axonal degeneration in the optic nerve in vivo. *Proc Natl Acad Sci USA*. 2010;107(13):6064-6069. doi:10.1073/pnas.0909794107
21. Gu X, Truong T, Heaster-Ford T, et al. Evaluating the Optic Nerve Crush Model to Understand the Function of Microglia in Glaucoma Neuroprotection. *Invest Ophthalmol Vis Sci*. 2025;66(12):56. doi:10.1167/iovs.66.12.56
22. Liu T, Wang YX, Jonas JB, Hoang QV, Girard MJA, Wang X. Gaze-Induced Optic Nerve Head Deformations Are Greater in High Myopia and Strains Increase With Axial Length. *Investigative Ophthalmology & Visual Science*. 2025;66(9):21-21. doi:10.1167/iovs.66.9.21
23. Wang X, Fisher LK, Milea D, Jonas JB, Girard MJA. Predictions of Optic Nerve Traction Forces and Peripapillary Tissue Stresses Following Horizontal Eye Movements. *Invest Ophthalmol Vis Sci*. 2017;58(4):2044. doi:10.1167/iovs.16-21319
24. Shen L, You QS, Xu X, et al. Scleral Thickness in Chinese Eyes. *Invest Ophthalmol Vis Sci*. 2015;56(4):2720. doi:10.1167/iovs.14-15631
25. Goyal V, Read AT, Brown DM, et al. Morphometric Analysis of Retinal Ganglion Cell Axons in Normal and Glaucomatous Brown Norway Rats Optic Nerves. *Trans Vis Sci Tech*. 2023;12(3):8. doi:10.1167/tvst.12.3.8
26. Mikelberg FS, Drance SM, Schulzer M, Yidegiligne HM, Weis MM. The Normal Human Optic Nerve. *Ophthalmology*. 1989;96(9):1325-1328. doi:10.1016/S0161-6420(89)32718-7
27. Jonas JB, Müller-Bergh JA, Schlötzer-Schrehardt UM, Naumann GO. Histomorphometry of the human optic nerve. *Investigative Ophthalmology & Visual Science*. 1990;31(4):736-744.
28. Cercignani M, Giulietti G, Dowell NG, et al. Characterizing axonal myelination within the healthy population: a tract-by-tract mapping of effects of age and gender on the fiber g-ratio. *Neurobiology of Aging*. 2017;49:109-118. doi:10.1016/j.neurobiolaging.2016.09.016

29. Geuzaine C, Remacle JF. Gmsh: A 3-D Finite Element Mesh Generator with Built-in Pre- and Post-Processing Facilities. *International Journal for Numerical Methods in Engineering*. 2009;79:1309-1331. doi:10.1002/nme.2579
30. Arbogast KB, Margulies SS. A fiber-reinforced composite model of the viscoelastic behavior of the brainstem in shear. *Journal of Biomechanics*. 1999;32(8):865-870. doi:10.1016/S0021-9290(99)00042-1
31. Hollows FC, Graham PA. Intra-ocular pressure, glaucoma, and glaucoma suspects in a defined population. *British Journal of Ophthalmology*. 1966;50(10):570-586. doi:10.1136/bjo.50.10.570
32. Leske MC, Heijl A, Hyman L, Bengtsson B. Early manifest glaucoma trial. *Ophthalmology*. 1999;106(11):2144-2153. doi:10.1016/S0161-6420(99)90497-9
33. Wang N, Xie X, Yang D, et al. Orbital Cerebrospinal Fluid Space in Glaucoma: The Beijing Intracranial and Intraocular Pressure (iCOP) Study. *Ophthalmology*. 2012;119(10):2065-2073.e1. doi:10.1016/j.ophtha.2012.03.054
34. Liu C, Reina C. Discrete Averaging Relations for Micro to Macro Transition. *Journal of Applied Mechanics*. 2016;83(081006). doi:10.1115/1.4033552
35. Maas SA, Ellis BJ, Ateshian GA, Weiss JA. FEBio: Finite Elements for Biomechanics. *Journal of Biomechanical Engineering*. 2012;134(1):011005. doi:10.1115/1.4005694
36. Kevenaar JT, Hoogenraad CC. The axonal cytoskeleton: from organization to function. *Front Mol Neurosci*. 2015;8. doi:10.3389/fnmol.2015.00044
37. Tang-Schomer MD, Patel AR, Baas PW, Smith DH. Mechanical breaking of microtubules in axons during dynamic stretch injury underlies delayed elasticity, microtubule disassembly, and axon degeneration. *The FASEB Journal*. 2010;24(5):1401-1410. doi:10.1096/fj.09-142844
38. Yu DY, Cringle SJ, Balaratnasingam C, Morgan WH, Yu PK, Su EN. Retinal ganglion cells: Energetics, compartmentation, axonal transport, cytoskeletons and vulnerability. *Progress in Retinal and Eye Research*. 2013;36:217-246. doi:10.1016/j.preteyeres.2013.07.001

39. Bansal M, Wang B, Waxman S, et al. Proposing a Methodology for Axon-Centric Analysis of IOP-Induced Mechanical Insult. *Invest Ophthalmol Vis Sci*. 2024;65(13):1. doi:10.1167/iovs.65.13.1
40. Quigley HA, Anderson DR. Distribution of axonal transport blockade by acute intraocular pressure elevation in the primate optic nerve head. *Investigative Ophthalmology & Visual Science*. 1977;16(7):640-644.
41. Chidlow G, Ebneter A, Wood JPM, Casson RJ. The optic nerve head is the site of axonal transport disruption, axonal cytoskeleton damage and putative axonal regeneration failure in a rat model of glaucoma. *Acta Neuropathol*. 2011;121(6):737-751. doi:10.1007/s00401-011-0807-1
42. Chan DD, Knutsen AK, Lu YC, et al. Statistical Characterization of Human Brain Deformation During Mild Angular Acceleration Measured In Vivo by Tagged Magnetic Resonance Imaging. *Journal of Biomechanical Engineering*. 2018;140(10):101005. doi:10.1115/1.4040230
43. Korneva A, Schaub J, Jefferys J, et al. A method to quantify regional axonal transport blockade at the optic nerve head after short term intraocular pressure elevation in mice. *Experimental Eye Research*. 2020;196:108035. doi:10.1016/j.exer.2020.108035
44. Chaudhary P, Stowell C, Reynaud J, et al. Optic Nerve Head Myelin-Related Protein, GFAP, and Iba1 Alterations in Non-Human Primates With Early to Moderate Experimental Glaucoma. *Invest Ophthalmol Vis Sci*. 2022;63(11):9. doi:10.1167/iovs.63.11.9
45. Chaudhary P, Lockwood H, Stowell C, et al. Retrolaminar Demyelination of Structurally Intact Axons in Nonhuman Primate Experimental Glaucoma. *Invest Ophthalmol Vis Sci*. 2024;65(2):36. doi:10.1167/iovs.65.2.36
46. Monavarfeshani A, Yan W, Pappas C, et al. Transcriptomic analysis of the ocular posterior segment completes a cell atlas of the human eye. *Proc Natl Acad Sci USA*. 2023;120(34):e2306153120. doi:10.1073/pnas.2306153120
47. Girard MJA, Suh JKF, Bottlang M, Burgoyne CF, Downs JC. Scleral Biomechanics in the Aging Monkey Eye. *Invest Ophthalmol Vis Sci*. 2009;50(11):5226. doi:10.1167/iovs.08-3363

48. Friberg TR, Lace JW. A comparison of the elastic properties of human choroid and sclera. *Experimental Eye Research*. 1988;47(3):429-436. doi:10.1016/0014-4835(88)90053-X
49. Chan WH, Hussain AA, Marshall J. Youngs Modulus of Bruchs Membrane: Implications for AMD. *Investigative Ophthalmology & Visual Science*. 2007;48(13):2187-2187.
50. Zhang L, Beotra MR, Baskaran M, et al. In Vivo Measurements of Prelamina and Lamina Cribrosa Biomechanical Properties in Humans. *Invest Ophthalmol Vis Sci*. 2020;61(3):27. doi:10.1167/iovs.61.3.27
51. Zhang L, Albon J, Jones H, et al. Collagen Microstructural Factors Influencing Optic Nerve Head Biomechanics. *Investigative Ophthalmology & Visual Science*. 2015;56(3):2031-2042. doi:10.1167/iovs.14-15734
52. Park J, Shin A, Jafari S, Demer JL. Material properties and effect of preconditioning of human sclera, optic nerve, and optic nerve sheath. *Biomech Model Mechanobiol*. 2021;20(4):1353-1363. doi:10.1007/s10237-021-01448-2
53. Jafari S, Hollister J, Kavehpour P, Demer JL. Shear viscoelastic properties of human orbital fat. *Journal of Biomechanics*. 2024;177:112416. doi:10.1016/j.jbiomech.2024.112416
54. Schoemaker I, Hoefnagel PPW, Mastenbroek TJ, et al. Elasticity, Viscosity, and Deformation of Orbital Fat. *Invest Ophthalmol Vis Sci*. 2006;47(11):4819. doi:10.1167/iovs.05-1497

# FIGURES

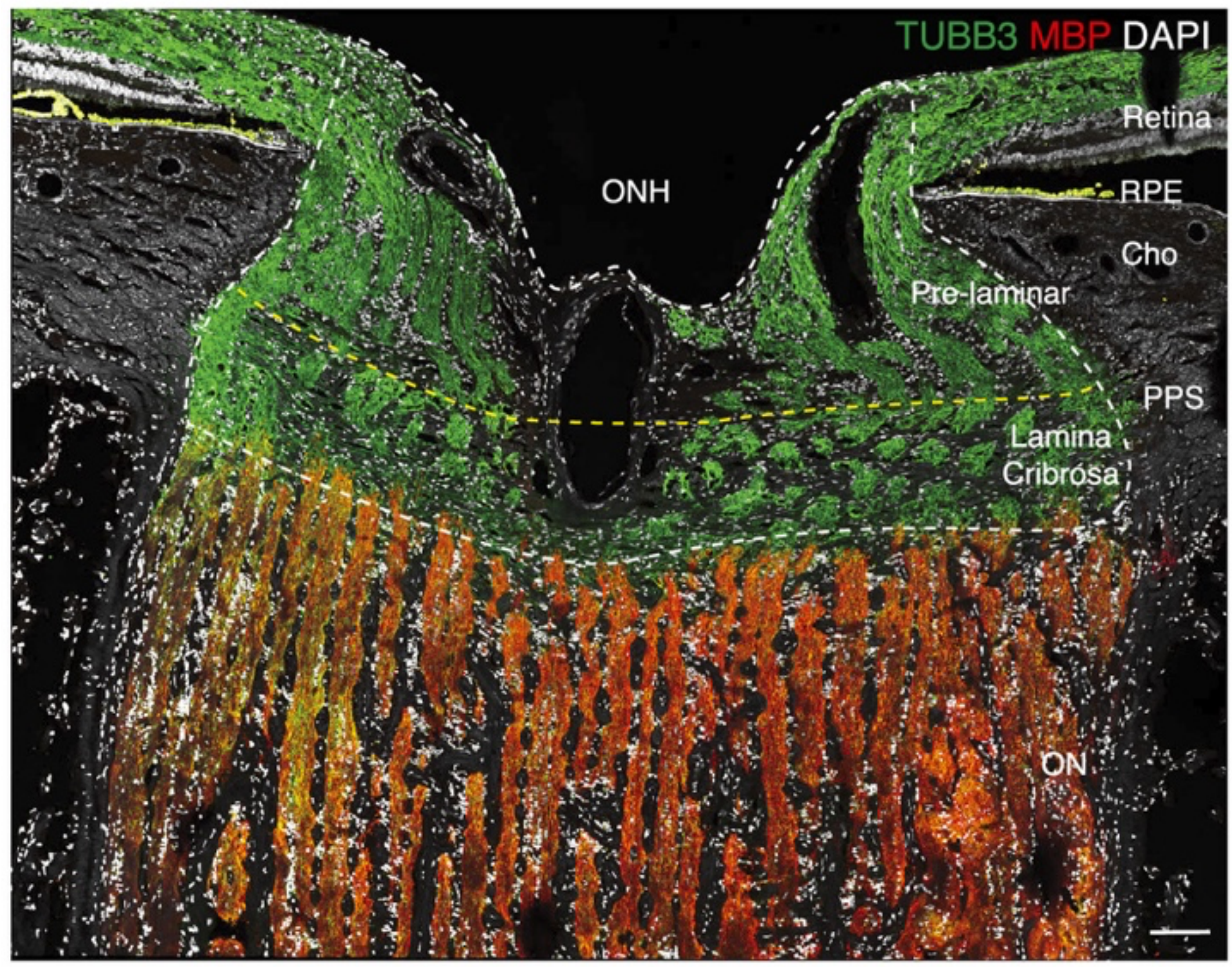

**Figure 1.** Section of the optic nerve head (ONH) and surrounding tissues immunostained for myelin basic protein (MBP, red) and beta-tubulin (TUBB3, green). TUBB3 highlights bundles of axons in the retina and ONH, and MBP highlights myelinating oligodendrocytes in the optic nerve (ON). Scale bar: 100 µm. PPS, peripapillary sclera; RPE, retinal pigment epithelium; Cho, choroid. [46] Source: Monavarfeshani et al., PNAS (2023). DOI: 10.1073/pnas.2306153120.

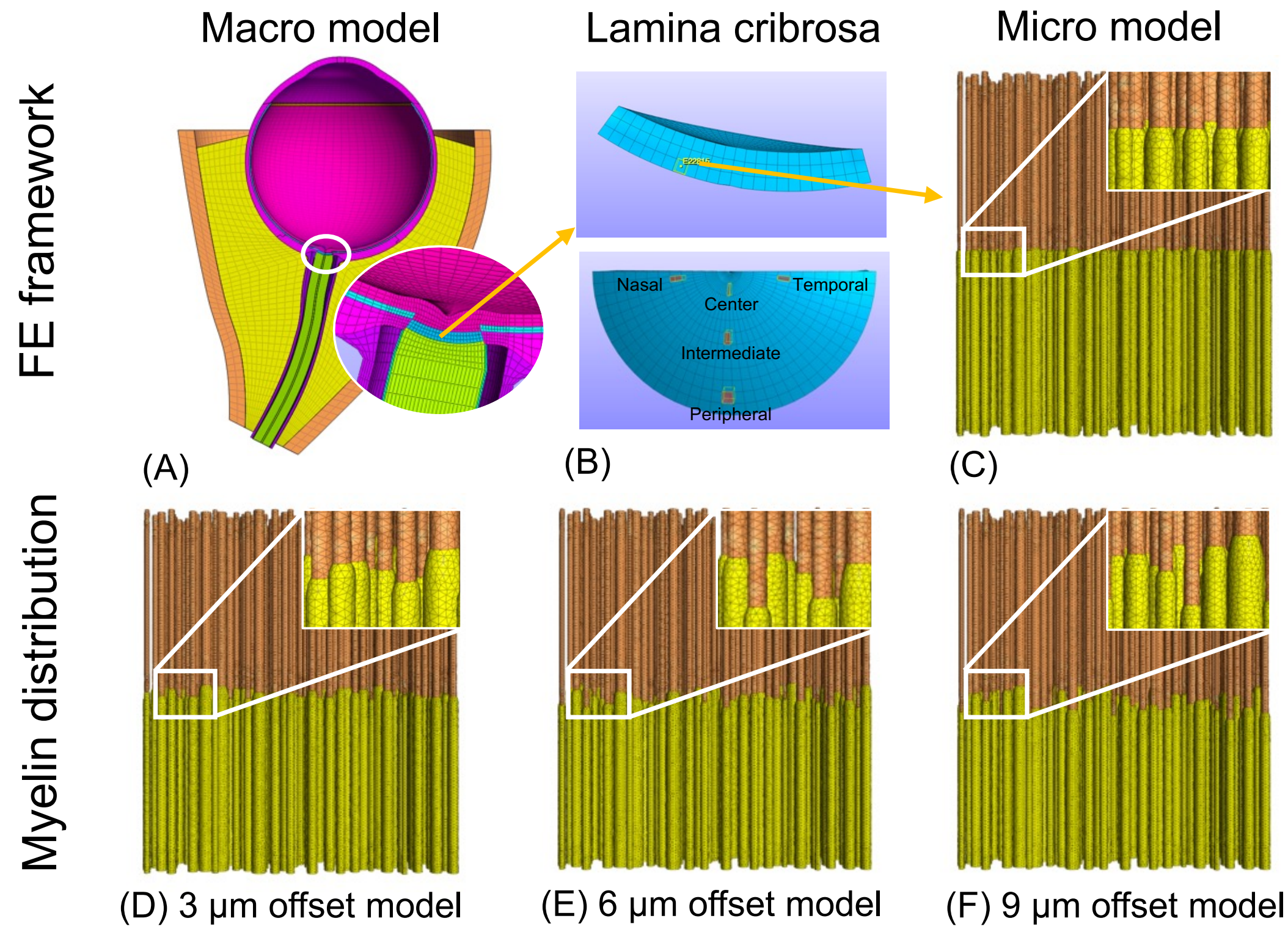

**Figure 2.** Overview of the multiscale modeling framework and myelin distribution patterns. (A) Macro-scale eye model with a magnified view of the optic nerve head region. (B) Lamina cribrosa highlighting the selected elements used for deformation extraction. (C) Micro-scale baseline model with a flat boundary defining the myelin transition zone. Brown shows unmyelinated axon segments; yellow shows myelin. (D–F) Myelin distribution patterns with posterior offsets randomly distributed within 0-3, 0-6, and 0-9 μm relative to the reference MTZ boundary, respectively. The values of 3, 6, and 9 μm represent the maximum prescribed offsets.

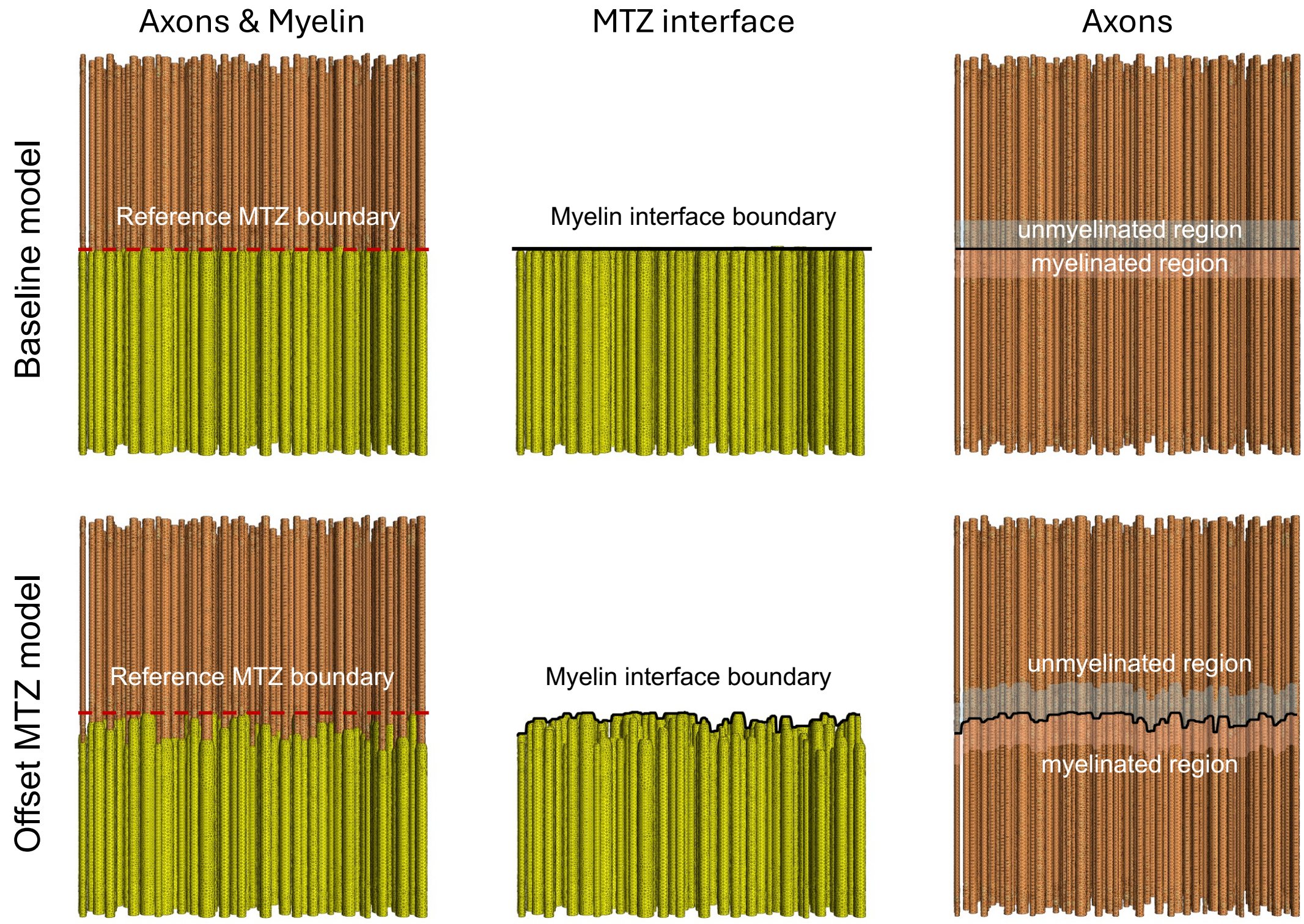

**Figure 3.** Definition of regions in the baseline and offset MTZ models. The red dashed line marks the reference MTZ boundary. The black line represents the myelin interface boundary. Two axonal regions relative to each axon's myelin interface are shown: blue represents an unmyelinated region extending from the myelin interface boundary 3 μm anterior (toward the lamina cribrosa), and pink represents a myelinated region extending 3 μm posterior (toward the retrolaminar optic nerve).

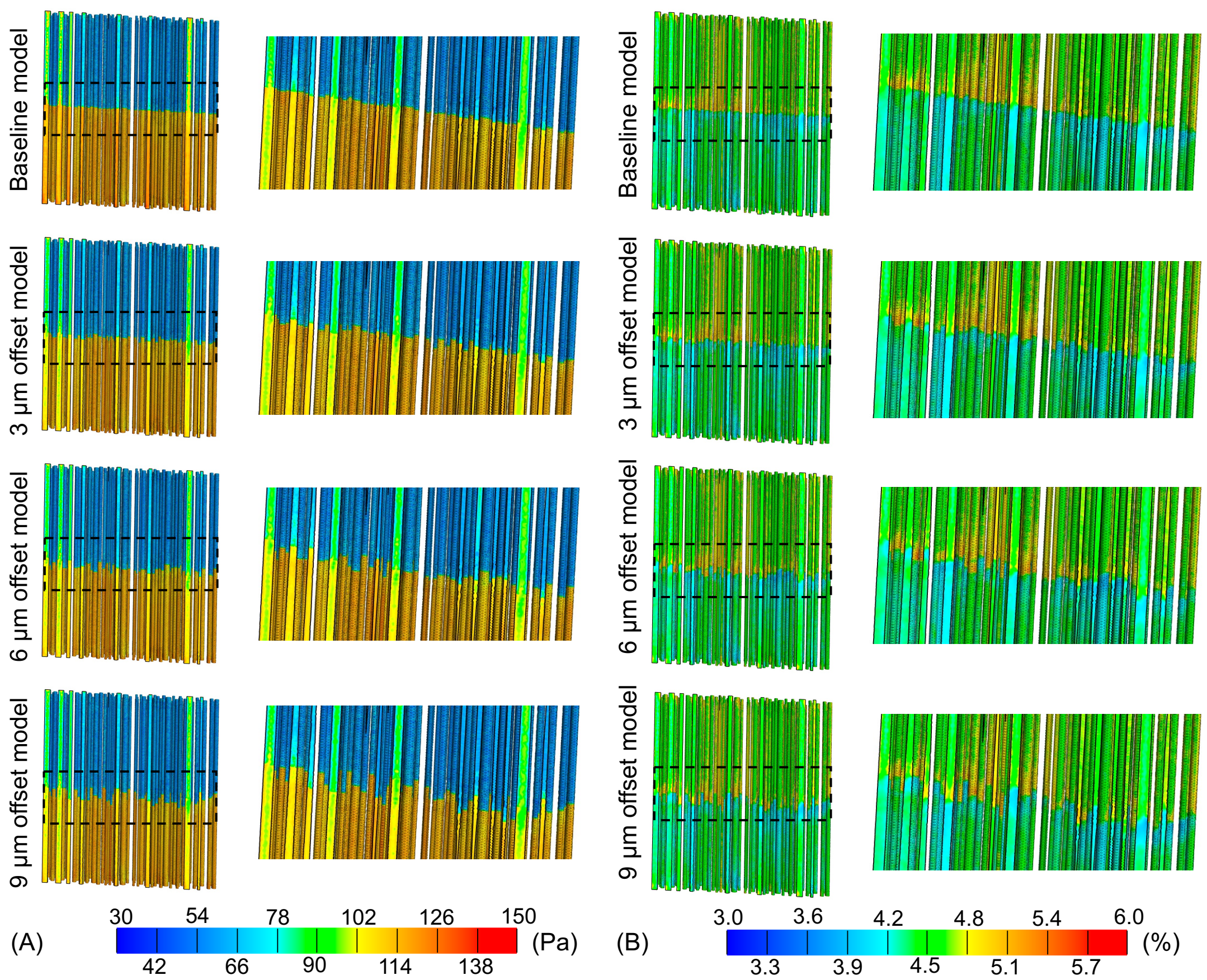

**Figure 4.** Effective stress (A) and strain (B) of axons under the elevated IOP condition. For each of the four MTZ configuration models (rows), the left panel displays the effective stress distribution in the full axon, while the right panel shows a 3× magnified view of the region within the dashed box around the MTZ.

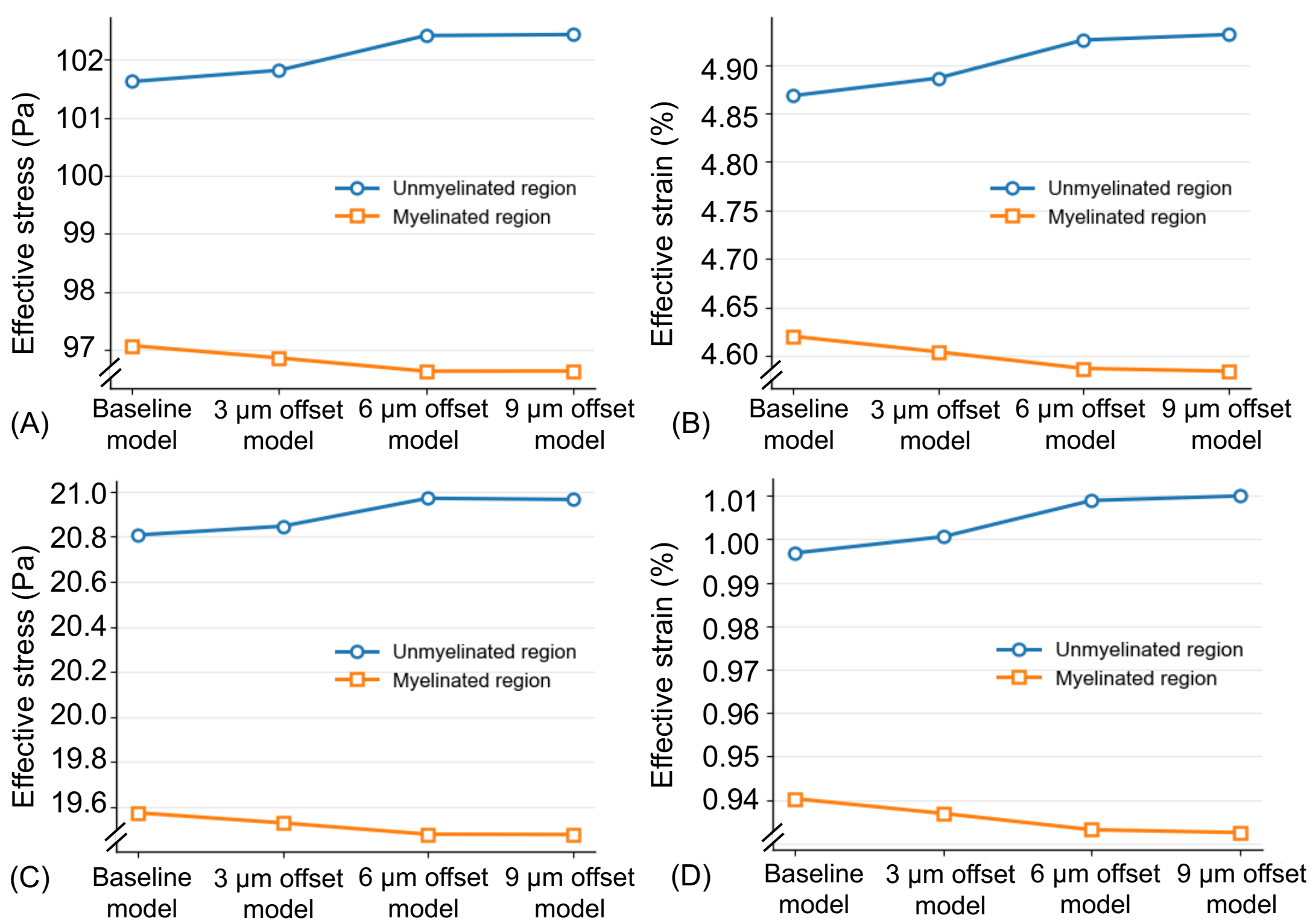

**Figure 5.** Effective stress and strain in the unmyelinated and myelinated regions across MTZ models under elevated IOP (A, B) and normal (C, D) IOP conditions.

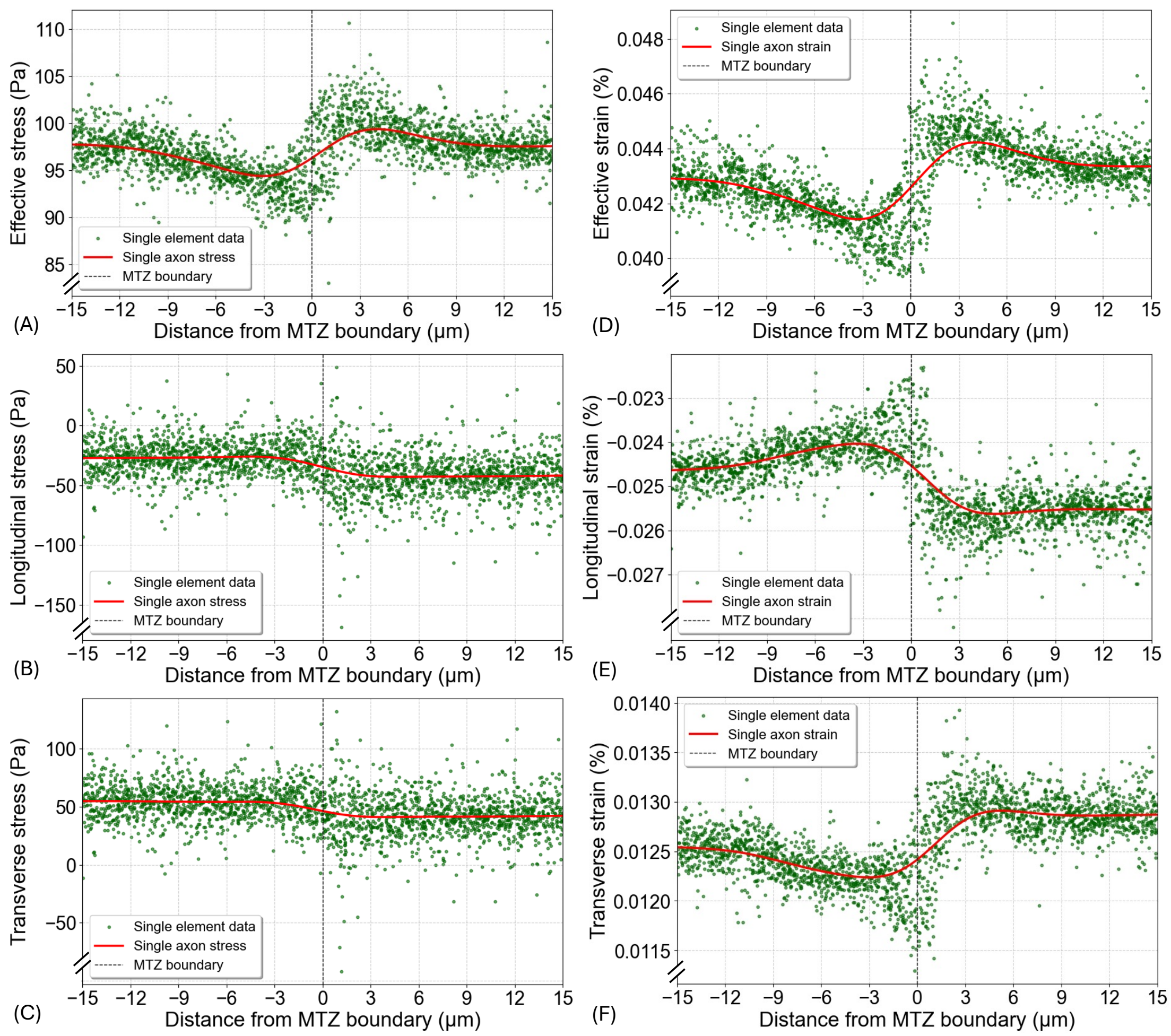

**Figure 6.** Single-axon stress and strain profiles across the MTZ boundary in the baseline model under elevated IOP. Green dots represent stress/strain values of individual axonal elements, and the red line represents the fitted curve based on these values. The dashed line indicates the MTZ boundary (0 μm) between the below myelinated (negative distance) and above unmyelinated (positive distance) regions. (A) Effective stress decreases locally at the boundary before increasing within the unmyelinated region. (B–C) Longitudinal stress increases toward the unmyelinated region, while transverse stress exhibits an opposite trend. (D–F) Strain components demonstrate similar boundary-associated variations across the MTZ.

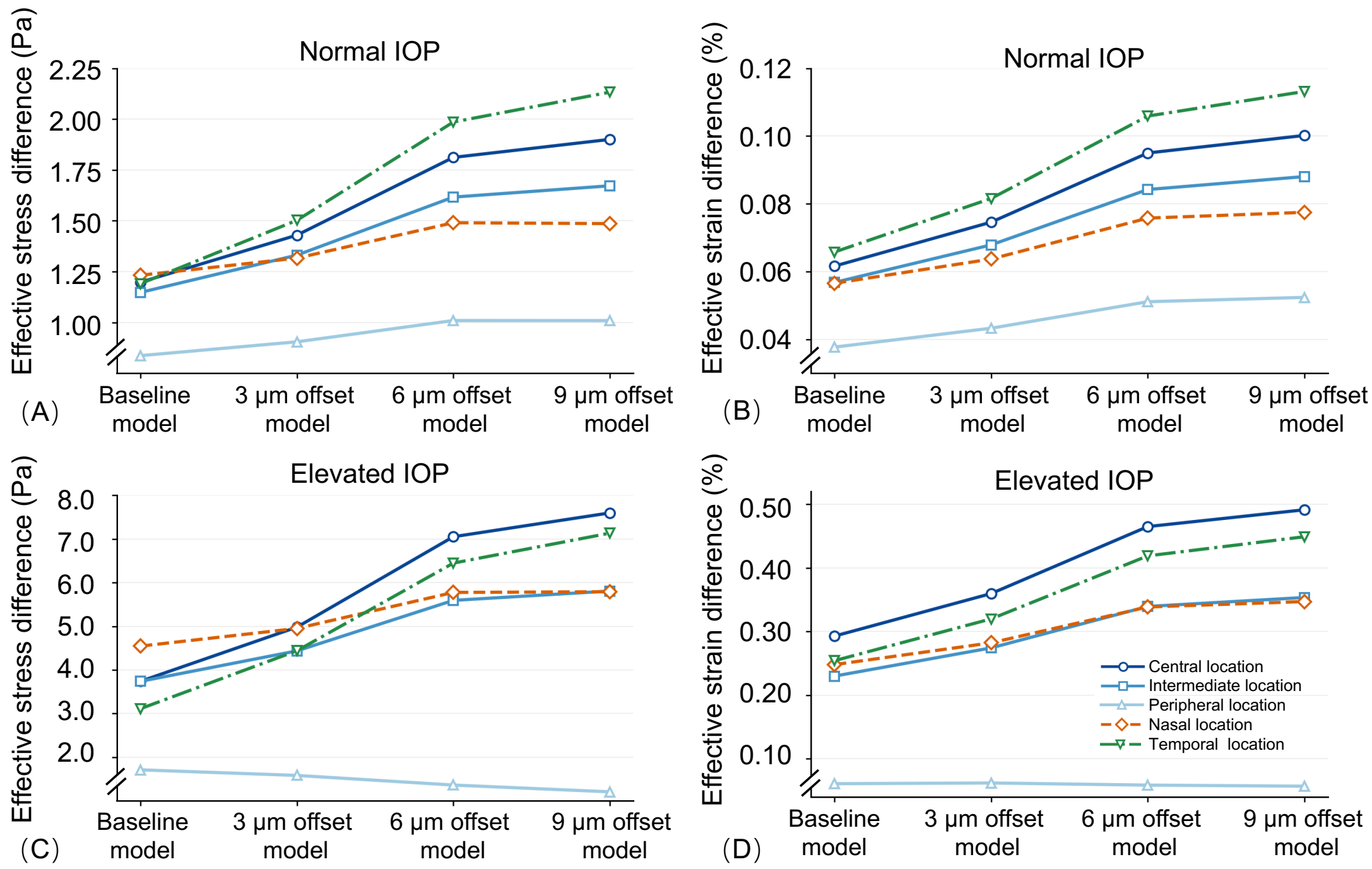

**Figure 7.** Quantitative comparison of axonal stress and strain difference. Effective stress (A, C) and effective strain (B, D) differences between regions extending 3 μm above and 3 μm below the myelin boundary under normal and elevated IOP conditions in five locations. The radial locations are shown using solid lines in the same blue color family, whereas the nasal (orange line) and temporal (green line) locations are shown using contrasting dashed lines.

# TABLES

**Table 1.** Tissue biomechanical properties in macro and micro model

| Tissue | Constitutive Model | Biomechanical Properties | References |
|---|---|---|---|
| Sclera | Mooney-Rivlin Von Mises Distributed Fibers | c1 = 0.805 MPa<br>c3 = 0.0127 MPa<br>c4 = 1102.25<br>kf = 2 (scleral ring)<br>kf = 0 (other region of sclera)<br>θp: preferred fiber orientations* | Girard et al. [47] |
| Choroid | Isotropic Elastic | Elastic modulus = 0.6 MPa<br>Poisson's ratio = 0.49 | Friberg et al. [48] |
| Bruch's membrane | Isotropic Elastic | Elastic modulus = 10.79MPa<br>Poisson's ratio = 0.49 | Chan et al. [49] |
| Retina | Isotropic Elastic | Elastic modulus = 0.1917MPa<br>Poisson's ratio = 0.49 | Zhang et al. [50] |
| Lamina Cribrosa | Mooney-Rivlin Von Mises Distributed Fibers | c1 = 0.05 MPa<br>c3 = 0.0025 MPa<br>c4 = 100<br>kf = 1<br>θp: preferred fiber orientation§ | Zhang et al. [51] |
| Optic nerve | Isotropic Elastic | Elastic modulus = 0.798MPa<br>Poisson's ratio = 0.49 | Park et al. [52] |
| Pia | Isotropic Elastic | Elastic modulus = 1.467 MPa<br>Poisson's ratio = 0.49 | Park et al. [52] |
| Dura | Isotropic Elastic | Elastic modulus = 1.467 MPa<br>Poisson's ratio = 0.49 | Park et al. [52] |
| OFM | Isotropic Elastic | Elastic modulus = 0.0022 MPa<br>Poisson's ratio = 0.49 | Jafari et al. [53] |
| Orbital bone | Isotropic Elastic | Elastic modulus = 300 MPa<br>Poisson's ratio = 0.49 | Schoemaker et al. [54] |
| Axon | Isotropic Elastic | Elastic modulus = 3411 Pa<br>Poisson's ratio = 0.49 | Weickenmeier et al. [5]<br>Chuang et al. [6]<br>Arbogast et al. [30] |
| Myelin | Isotropic Elastic | Elastic modulus = 4154 Pa<br>Poisson's ratio = 0.49 | |
| Matrix | Isotropic Elastic | Elastic modulus = 1137 Pa<br>Poisson's ratio = 0.49 | |

*Collagen fibers in the scleral ring were aligned circumferentially around the scleral canal. Fibers in other parts of the sclera were organized randomly.
§Collagen fibers in the LC were organized in the radial direction (from the central vessel trunk to the scleral canal)

# Supplementary Material A

## Mesh analysis of the micro-model

A mesh refinement/coarsening analysis was performed to evaluate the sensitivity of the regional outcomes to mesh resolution. The same micro-model (baseline model: flat myelin boundary) was used to generate conforming tetrahedral elements automatically with the Gmsh Python API. Four prescribed mesh sizes of 0.9, 0.8, 0.7, and 0.6 μm were evaluated. This range was selected to balance accurate representation of the smallest geometric features against the computational cost associated with further mesh refinement. The geometry, material properties, macro-model deformation input, boundary conditions, loading conditions, and solver settings were maintained across all mesh densities. Mesh sensitivity was assessed by comparing the mean effective stress and effective strain within the predefined 3-μm myelinated and unmyelinated regions.

The regional mean effective stress and strain values remained stable with progressive mesh refinement (**Table S1**). Relative to the refined 0.6-μm mesh, the 0.8-μm mesh used in the main analyses differed by only 0.56% and 0.73% in terms of the mean effective stress in the myelinated and unmyelinated regions, respectively. The corresponding differences in mean effective strain were 0.40% and 0.73%. Across all evaluated mesh densities, the maximum relative difference in any regional mean outcome was less than 1%. Moreover, the unmyelinated region consistently exhibited greater mean effective stress and strain than the myelinated region. These results indicate that the regional outcomes and the associated conclusions were not unduly influenced by the selected mesh resolution.

**Table S1.** Mesh refinement/coarsening analysis of regional mean effective stress and strain in the micro model.

| Mesh size (μm) | Nodes | Elements | Myelinated effective stress (Pa) | Unmyelinated effective stress (Pa) | Myelinated effective strain (%) | Unmyelinated effective stress (%) |
|---|---|---|---|---|---|---|
| 0.6 | 1,055,131 | 6,252,280 | 87.22 | 93.18 | 5.00 | 5.45 |
| 0.7 | 711,115 | 4,204,502 | 87.47 | 92.79 | 5.01 | 5.43 |
| **0.8** | 506,967 | 2,984,041 | 87.70 | 92.50 | 5.02 | 5.41 |
| 0.9 | 375,438 | 2,204,311 | 87.79 | 92.26 | 5.03 | 5.40 |

# Supplementary Material B

## Macro-scale locations and deformation gradient tensors

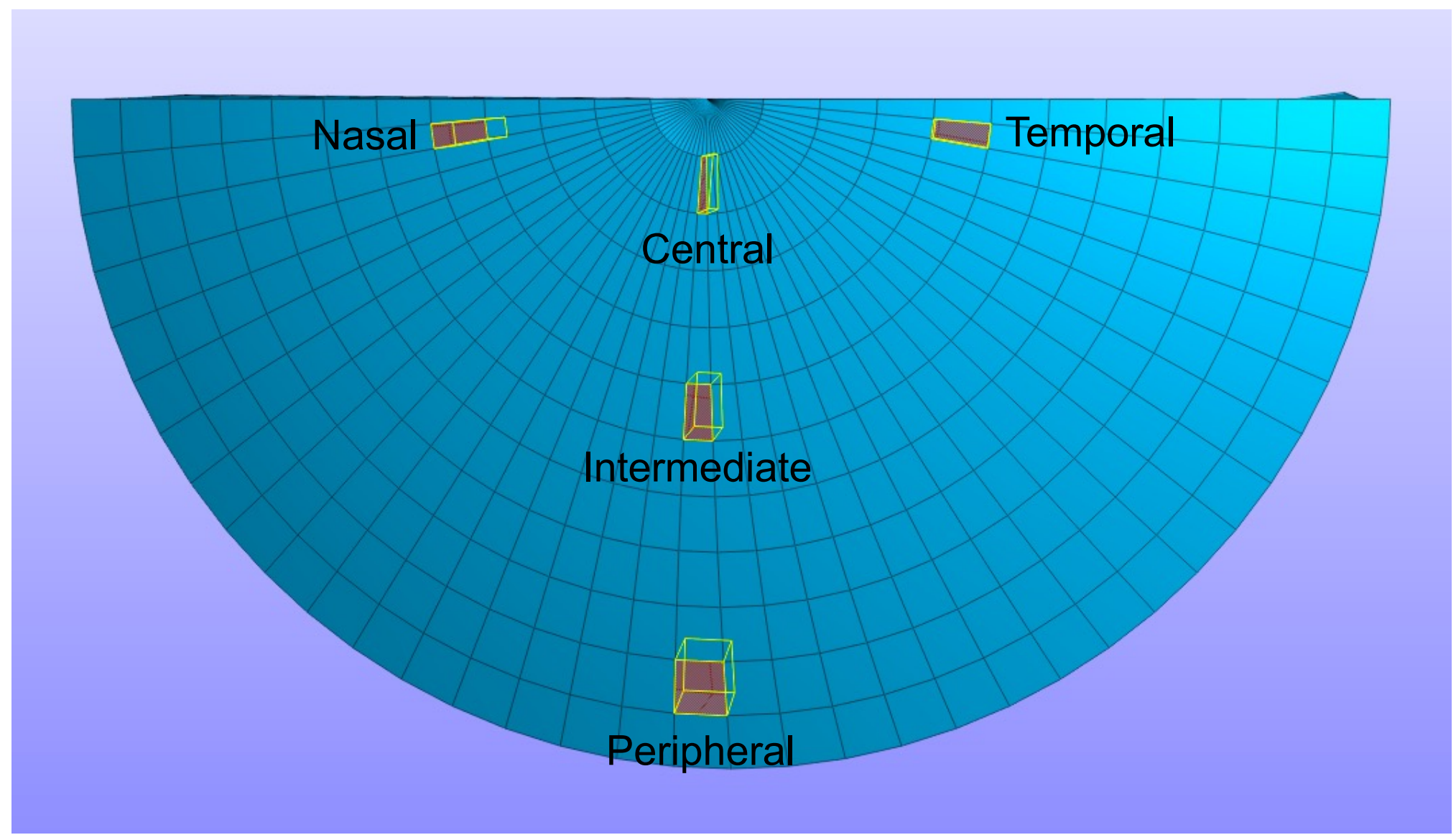

**Figure S1.** Locations of the five macro-scale elements used in the spatial sensitivity analysis. Three elements were selected along a common radial transect, representing LC central, intermediate, and peripheral locations. Two additional elements located at comparable radial positions represented the nasal and temporal regions to evaluate circumferential variation. Location-specific deformation gradients were applied separately to the same micro-scale geometry under both normal and elevated IOP conditions.

The average effective strain of the entire LC was 0.82% and 3.09% under normal and elevated IOP, respectively. The deformation gradient tensor at each of the five selected macro-scale locations are summarized in **Table S2**. All deformation gradient tensors were expressed in a Cartesian coordinate system, with the y-axis aligned with the longitudinal direction of the axons and the x- and z-axes representing the transverse directions.

**Table S2. The deformation gradient tensors of five macro-scale elements under normal and elevated IOP**

| Location | $\boldsymbol{F}$, normal IOP | $\boldsymbol{F}$， elevated IOP |
|---|---|---|
| Nasal region | $\begin{bmatrix} 1.0024 & 3.78 \times 10^{-3} & -1.35 \times 10^{-4} \\ -1.01 \times 10^{-2} & 0.9948 & -6.00 \times 10^{-4} \\ -9.09 \times 10^{-6} & 1.60 \times 10^{-4} & 1.0028 \end{bmatrix}$ | $\begin{bmatrix} 1.0100 & 1.57 \times 10^{-2} & -1.16 \times 10^{-3} \\ -4.94 \times 10^{-2} & 0.9744 & -5.20 \times 10^{-3} \\ 1.56 \times 10^{-4} & 2.59 \times 10^{-3} & 1.0152 \end{bmatrix}$ |
| Temporal region | $\begin{bmatrix} 1.0046 & 1.20 \times 10^{-3} & 4.78 \times 10^{-5} \\ -2.01 \times 10^{-3} & 0.9920 & -5.36 \times 10^{-3} \\ 9.87 \times 10^{-6} & 1.74 \times 10^{-4} & 1.0034 \end{bmatrix}$ | $\begin{bmatrix} 1.0180 & -2.29 \times 10^{-3} & 7.60 \times 10^{-5} \\ 2.49 \times 10^{-2} & 0.9663 & -4.84 \times 10^{-3} \\ 7.50 \times 10^{-4} & 2.51 \times 10^{-3} & 1.0159 \end{bmatrix}$ |
| LC Central region | $\begin{bmatrix} 1.0040 & 2.88 \times 10^{-3} & -1.72 \times 10^{-4} \\ -7.08 \times 10^{-3} & 0.9929 & -1.13 \times 10^{-3} \\ -7.97 \times 10^{-5} & 2.45 \times 10^{-4} & 1.0031 \end{bmatrix}$ | $\begin{bmatrix} 1.0204 & -5.40 \times 10^{-3} & -2.35 \times 10^{-3} \\ -1.60 \times 10^{-2} & 0.9633 & -4.84 \times 10^{-3} \\ 8.63 \times 10^{-4} & 6.29 \times 10^{-3} & 1.0715 \end{bmatrix}$ |
| Model intermediate region | $\begin{bmatrix} 1.0035 & 2.81 \times 10^{-3} & -1.02 \times 10^{-3} \\ -6.73 \times 10^{-3} & 0.9937 & -5.36 \times 10^{-3} \\ -5.04 \times 10^{-5} & 1.78 \times 10^{-3} & 1.0028 \end{bmatrix}$ | $\begin{bmatrix} 1.0150 & -1.92 \times 10^{-3} & -8.76 \times 10^{-3} \\ -1.45 \times 10^{-2} & 0.9736 & -4.54 \times 10^{-2} \\ 3.79 \times 10^{-3} & 2.38 \times 10^{-2} & 1.0106 \end{bmatrix}$ |
| LC periphery region | $\begin{bmatrix} 1.0026 & 2.29 \times 10^{-3} & -2.16 \times 10^{-3} \\ -5.09 \times 10^{-3} & 0.9963 & -9.06 \times 10^{-3} \\ 4.51 \times 10^{-4} & 4.81 \times 10^{-3} & 1.0011 \end{bmatrix}$ | $\begin{bmatrix} 1.0065 & 3.87 \times 10^{-3} & -1.02 \times 10^{-2} \\ -7.25 \times 10^{-3} & 0.9992 & -4.86 \times 10^{-2} \\ 5.20 \times 10^{-3} & 3.27 \times 10^{-2} & 0.9925 \end{bmatrix}$ |

## Supplementary Material C

**Table S3. Detailed quantitative results of unmyelinated and myelinated regions**

| ***Nasal region*** | | Effective stress (Pa) | | | Effective strain (%) | | |
|---|---|---|---|---|---|---|---|
| | Model | Unmyelinated region | Myelinated region | Difference | Unmyelinated region | Myelinated region | Difference |
| Elevated IOP condition | Baseline | 101.62 | 97.07 | 4.55 | 4.87 | 4.62 | 0.25 |
| | 3 μm offset | 101.81 | 96.86 | 4.95 | 4.89 | 4.60 | 0.28 |
| | 6 μm offset | 102.41 | 96.63 | 5.78 | 4.93 | 4.59 | 0.34 |
| | 9 μm offset | 102.43 | 96.64 | 5.79 | 4.93 | 4.58 | 0.35 |
| Normal IOP condition | Baseline | 20.81 | 19.58 | 1.23 | 1.00 | 0.94 | 0.06 |
| | 3 μm offset | 20.85 | 19.53 | 1.32 | 1.00 | 0.94 | 0.06 |
| | 6 μm offset | 20.97 | 19.48 | 1.49 | 1.01 | 0.93 | 0.08 |
| | 9 μm offset | 20.96 | 19.48 | 1.49 | 1.01 | 0.93 | 0.08 |

| ***Temporal region*** | | Effective stress (Pa) | | | Effective strain (%) | | |
|---|---|---|---|---|---|---|---|
| | Model | Unmyelinated region | Myelinated region | Difference | Unmyelinated region | Myelinated region | Difference |
| Elevated IOP condition | Baseline | 113.31 | 110.20 | 3.11 | 5.84 | 5.59 | 0.25 |
| | 3 μm offset | 114.05 | 109.61 | 4.44 | 5.88 | 5.56 | 0.32 |
| | 6 μm offset | 115.40 | 108.95 | 6.45 | 5.95 | 5.53 | 0.42 |
| | 9 μm offset | 115.86 | 108.72 | 7.13 | 5.97 | 5.52 | 0.45 |
| Normal IOP condition | Baseline | 27.53 | 26.34 | 1.19 | 1.40 | 1.33 | 0.07 |
| | 3 μm offset | 27.71 | 26.21 | 1.50 | 1.41 | 1.33 | 0.08 |
| | 6 μm offset | 28.04 | 26.05 | 1.99 | 1.43 | 1.32 | 0.11 |
| | 9 μm offset | 28.14 | 26.01 | 2.13 | 1.43 | 1.32 | 0.11 |

| *Center region* | | Effective stress (Pa) | | | Effective strain (%) | | |
|---|---|---|---|---|---|---|---|
| | Model | Unmyelinated region | Myelinated region | Difference | Unmyelinated region | Myelinated region | Difference |
| Elevated IOP condition | Baseline | 129.68 | 125.94 | 3.74 | 6.56 | 6.27 | 0.29 |
| | 3 μm offset | 130.37 | 125.39 | 4.98 | 6.60 | 6.24 | 0.36 |
| | 6 μm offset | 131.78 | 124.72 | 7.05 | 6.67 | 6.21 | 0.47 |
| | 9 μm offset | 132.15 | 124.55 | 7.60 | 6.69 | 6.20 | 0.49 |
| Normal IOP condition | Baseline | 25.15 | 23.95 | 1.20 | 1.26 | 1.19 | 0.06 |
| | 3 μm offset | 25.28 | 23.85 | 1.43 | 1.26 | 1.19 | 0.07 |
| | 6 μm offset | 25.55 | 23.74 | 1.81 | 1.28 | 1.18 | 0.10 |
| | 9 μm offset | 25.61 | 23.71 | 1.90 | 1.28 | 1.18 | 0.10 |

| ***Intermediate region*** | | Effective stress (Pa) | | | Effective strain (%) | | |
|---|---|---|---|---|---|---|---|
| | Model | Unmyelinated region | Myelinated region | Difference | Unmyelinated region | Myelinated region | Difference |
| Elevated IOP condition | Baseline | 98.17 | 94.43 | 3.74 | 4.82 | 4.59 | 0.23 |
| | 3 μm offset | 98.56 | 94.12 | 4.44 | 4.85 | 4.58 | 0.27 |
| | 6 μm offset | 99.38 | 93.78 | 5.59 | 4.89 | 4.56 | 0.34 |
| | 9 μm offset | 99.51 | 93.70 | 5.81 | 4.90 | 4.55 | 0.35 |
| Normal IOP condition | Baseline | 23.04 | 21.89 | 1.15 | 1.13 | 1.07 | 0.06 |
| | 3 μm offset | 23.15 | 21.82 | 1.33 | 1.14 | 1.07 | 0.07 |
| | 6 μm offset | 23.35 | 21.74 | 1.62 | 1.15 | 1.06 | 0.08 |
| | 9 μm offset | 23.39 | 21.71 | 1.67 | 1.15 | 1.06 | 0.09 |

| ***Peripheral region*** | | Effective stress (Pa) | | | Effective strain (%) | | |
|---|---|---|---|---|---|---|---|
| | Model | Unmyelinated region | Myelinated region | Difference | Unmyelinated region | Myelinated region | Difference |
| Elevated IOP condition | Baseline | 33.82 | 32.11 | 1.71 | 1.43 | 1.37 | 0.06 |
| | 3 µm offset | 33.77 | 32.19 | 1.58 | 1.43 | 1.37 | 0.06 |
| | 6 µm offset | 33.67 | 32.31 | 1.36 | 1.43 | 1.37 | 0.06 |
| | 9 µm offset | 33.57 | 32.36 | 1.21 | 1.43 | 1.37 | 0.06 |
| Normal IOP condition | Baseline | 15.42 | 14.58 | 0.84 | 0.72 | 0.68 | 0.04 |
| | 3 µm offset | 15.46 | 14.56 | 0.90 | 0.72 | 0.68 | 0.04 |
| | 6 µm offset | 15.55 | 14.54 | 1.01 | 0.73 | 0.68 | 0.05 |
| | 9 µm offset | 15.54 | 14.53 | 1.01 | 0.73 | 0.68 | 0.05 |

# Supplementary Material D

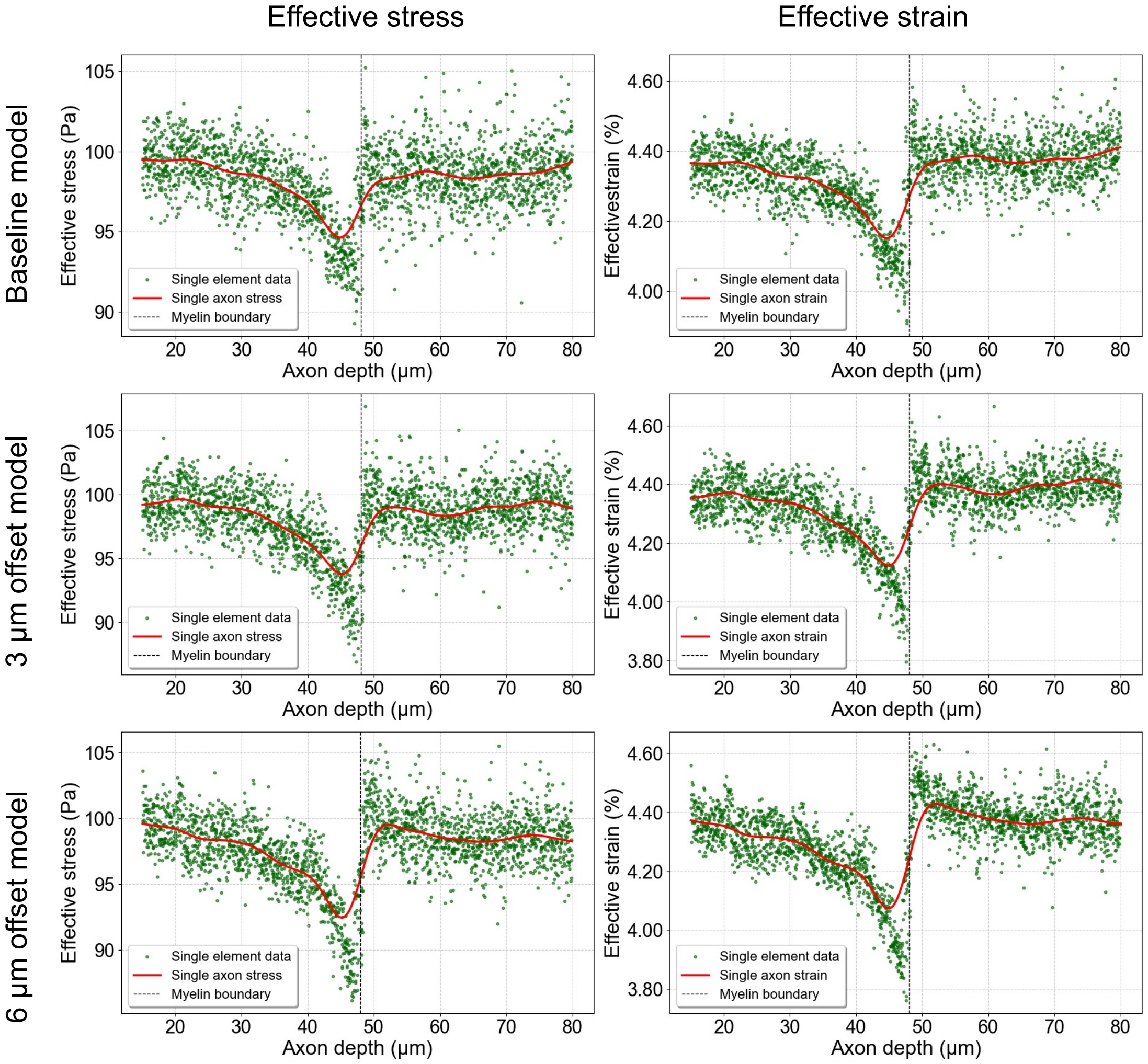

**Figure S2.** Comparison of effective stress and strain in selected axons across MTZ configuration models. Axons with identical radii, spatial locations, and nearly identical myelin heights (48.06, 48.01, and 48.00 μm) were used for cross-model comparison. Under these matched conditions, increasing myelin offset led to more pronounced stress and strain discontinuities at the myelin boundary.